\documentstyle[aps,prb,tighten]{revtex}
\begin{document}
\twocolumn[\hsize\textwidth\columnwidth\hsize
\csname@twocolumnfalse\endcsname

\title{Local properties in the two-dimensional $t$-$t^{\prime }$-$U$ model}
\author{Adolfo Avella, Ferdinando Mancini and Dario Villani}
\address{Dipartimento di Fisica Teorica e S.M.S.A. - Unit\`a INFM di Salerno\\
Universit\`a di Salerno, 84081 Baronissi (SA), Italy}
\author{Hideki Matsumoto}
\address{Institute for Materials Research, Tohoku University, 980 Sendai, Japan}
\date{1 September, 1996}
\maketitle

\begin{abstract}\widetext
We have studied the $t$-$t^{\prime }$-$U$ model by means of the composite
operator method. The effect of the bare diagonal hopping term $t^{\prime }$
that appears to be material dependent for high-$T_c$ cuprate superconductors
is analyzed in detail. In particular, some local quantities are computed and
a comprehensive comparison with the data by numerical simulations on finite
size lattices is presented. The result show a good agreement with those
obtained by Monte Carlo methods.
\end{abstract}
\pacs{}]

\narrowtext

\section{Introduction}

Since the discovery of the high-$T_c$ superconductivity, there has been a
great deal of discussion about the choice of an effective model suitable to
describe the properties of the copper-oxide planes in the perovskite
structure. Extensive studies of the magnetic properties, showing one spin
degree of freedom in the $Cu$-$O$ plane\cite{Monien:1991}, have resulted in
considerable evidence that the high-temperature superconductors may be
modelled by an effective single-band model. In this line of thinking, one of
the most studied model is the single-band Hubbard model which indeed can
qualitatively describe many physical properties experimentally observed in
copper-oxide compounds. On the other hand, a particle-hole symmetric model
cannot distinguish between electron- and hole-doped materials. The addition
of a finite diagonal hopping term $t^{\prime }$ has often been suggested to
handle the complexity of the experimental situation for the cuprate being
essential to reproduce various experimental observations. Moreover, this
electron-hole asymmetry in the next-nearest-neighbor hopping term, combined
with a perfect symmetry of all the other effective parameters, emerges from
various reduction procedures of multi-component electronic models and seems
to distinguish the cuprates from a general charge-transfer insulator\cite
{Feiner:1996}.

In the next section the formulas for the $t$-$t^{\prime }$-$U$ model in the
composite operator method $\left( COM\right) $\cite{Mancini:1995a} framework
are summarized. Our results and a comparison with data of numerical analysis
by quantum Monte Carlo method\cite{Duffy:1995} are also presented. Some
conclusions are given in Sec. \ref{con}.

\section{Results}

Let us consider the $t$-$t^{\prime }$-$U$ model described by the
Hamiltonian: 
\begin{eqnarray}
&&H=-\mu \sum_ic^{\dagger }\left( i\right) c\left( i\right)
-t\sum_{ij}\alpha _{ij}c^{\dagger }\left( i\right) c\left( j\right)  
\nonumber \\
&&-t^{\prime }\sum_{ij}\beta _{ij}c^{\dagger }\left( i\right) c\left(
j\right) +U\sum_in_{\uparrow }\left( i\right) n_{\downarrow }\left( i\right) 
\end{eqnarray}
where for a two-dimensional quadratic lattice with lattice constant $a$%
\begin{eqnarray}
&&\alpha \left( {\bf k}\right) =\frac 12\left( \cos \left( k_xa\right) +\cos
\left( k_ya\right) \right)  \\
&&\beta \left( {\bf k}\right) =\cos \left( k_xa\right) \cos \left(
k_ya\right) .
\end{eqnarray}
We use the spinor notation and drop the index of the spin freedom of
electrons unless when it is necessary, 
\begin{equation}
c=\left( 
\begin{array}{l}
c_{\uparrow } \\ 
c_{\downarrow }
\end{array}
\right) \qquad c^{\dagger }=\left( c_{\uparrow }^{\dagger }\quad
c_{\downarrow }^{\dagger }\right) .
\end{equation}

Following the $COM$\ ideas, we are interested in choosing a suitable
asymptotic field for new bound states which appear, due to the strong
correlations. Therefore, we introduce the doublet composite field operator 
\begin{equation}
\psi \left( i\right) =\left( 
\begin{array}{l}
\xi \left( i\right) \\ 
\eta \left( i\right)
\end{array}
\right)
\end{equation}
with 
\begin{eqnarray}
&&\xi _\sigma \left( i\right) =c_\sigma \left( i\right) \left( 1-n_{-\sigma
}\left( i\right) \right) \\
&&\eta _\sigma \left( i\right) =c_\sigma \left( i\right) n_{-\sigma }\left(
i\right) .
\end{eqnarray}

The properties of the system are conveniently expressed in terms of the
two-point retarded thermal Green function: 
\begin{equation}
S\left( i,j\right) =\left\langle R\left[ \psi \left( i\right) \psi ^{\dagger
}\left( j\right) \right] \right\rangle .
\end{equation}
In the static approximation\cite{Mancini:1995a}, the Fourier transform of $%
S\left( i,j\right) $ is given by 
\begin{equation}
S\left( {\bf k},\omega \right) =\frac 1{\omega -m\left( {\bf k}\right)
I^{-1}\left( {\bf k}\right) }I\left( {\bf k}\right)
\end{equation}
where $I\left( {\bf k}\right) $ and $m\left( {\bf k}\right) $ are defined as 
\begin{eqnarray}
&&I\left( {\bf k}\right) =\left\langle \left\{ \psi \left( i\right) ,\psi
^{\dagger }\left( j\right) \right\} \right\rangle _{F.T.} \\
&&m\left( {\bf k}\right) =\left\langle \left\{ i\frac \partial {\partial t}%
\psi \left( i\right) ,\psi ^{\dagger }\left( j\right) \right\} \right\rangle
_{F.T.}.
\end{eqnarray}

By considering a paramagnetic ground state, a straightforward calculation
gives 
\begin{eqnarray}
&&I\left( {\bf k}\right) =\left( 
\begin{array}{ll}
I_{11} & 0 \\ 
0 & I_{22}
\end{array}
\right) =\left( 
\begin{array}{ll}
1-\frac n2 & 0 \\ 
0 & \frac n2
\end{array}
\right)  \\
&&m_{11}\left( {\bf k}\right) =-\mu I_{11}-4t\left( \Delta +\alpha \left( 
{\bf k}\right) \left( p+1-n\right) \right)   \nonumber \\
&&-4t^{\prime }\left( \Delta ^{\prime }+\beta \left( {\bf k}\right) \left(
p^{\prime }+1-n\right) \right)  \\
&&m_{12}\left( {\bf k}\right) =m_{21}\left( {\bf k}\right) =4t\left( \Delta
+\alpha \left( {\bf k}\right) \left( I_{22}-p\right) \right)   \nonumber \\
&&+4t^{\prime }\left( \Delta ^{\prime }+\beta \left( {\bf k}\right) \left(
I_{22}-p^{\prime }\right) \right)  \\
&&m_{22}\left( {\bf k}\right) =\left( -\mu +U\right) I_{22}-4t\left( \Delta
+\alpha \left( {\bf k}\right) p\right)   \nonumber \\
&&-4t^{\prime }\left( \Delta ^{\prime }+\beta \left( {\bf k}\right)
p^{\prime }\right) .
\end{eqnarray}
We use the following notation 
\begin{eqnarray}
&&\psi ^\alpha \left( i\right) =\sum_j\alpha _{ij}\psi \left( j\right)  \\
&&\psi ^\beta \left( i\right) =\sum_j\beta _{ij}\psi \left( j\right)  \\
&&\Delta =\left\langle \xi ^\alpha \left( i\right) \xi ^{\dagger }\left(
i\right) \right\rangle -\left\langle \eta ^\alpha \left( i\right) \eta
^{\dagger }\left( i\right) \right\rangle  \\
&&\Delta ^{\prime }=\left\langle \xi ^\beta \left( i\right) \xi ^{\dagger
}\left( i\right) \right\rangle -\left\langle \eta ^\beta \left( i\right)
\eta ^{\dagger }\left( i\right) \right\rangle  \\
&&p=\frac 14\left\langle n_\mu ^\alpha \left( i\right) n_\mu \left( i\right)
\right\rangle -\left\langle c_{\uparrow }\left( i\right) c_{\downarrow
}\left( i\right) \left( c_{\downarrow }^{\dagger }\left( i\right)
c_{\uparrow }^{\dagger }\left( i\right) \right) ^\alpha \right\rangle  \\
&&p^{\prime }=\frac 14\left\langle n_\mu ^\beta \left( i\right) n_\mu \left(
i\right) \right\rangle -\left\langle c_{\uparrow }\left( i\right)
c_{\downarrow }\left( i\right) \left( c_{\downarrow }^{\dagger }\left(
i\right) c_{\uparrow }^{\dagger }\left( i\right) \right) ^\beta
\right\rangle 
\end{eqnarray}
$n_\mu \left( i\right) =c^{\dagger }\left( i\right) \sigma _\mu c\left(
i\right) $ being the charge $\left( \mu =0\right) $ and spin $\left( \mu
=1,2,3\right) $ density operator.

The quantities $\Delta $ and $\Delta ^{\prime }$ are self-consistent
parameters in the sense that they can be expressed in terms of the matrix
elements related to the fermion propagator. The parameters $p,$ $p^{\prime }$
and $\mu $, the chemical potential, can be fixed by self-consistent
equations 
\begin{eqnarray}
&&n=2\left( 1-\left\langle \xi \left( i\right) \xi ^{\dagger }\left(
i\right) \right\rangle -\left\langle \eta \left( i\right) \eta ^{\dagger
}\left( i\right) \right\rangle \right) \\
&&\left\langle \xi \left( i\right) \eta ^{\dagger }\left( i\right)
\right\rangle =0.
\end{eqnarray}
The details will be presented elsewhere. The solution of the set of
self-consistent equations allow us to compute the fermion Green function.

We have computed the chemical potential and the double occupancy $%
D=\left\langle n_{\uparrow }\left( i\right) n_{\downarrow }\left( i\right)
\right\rangle $ for different values of the particle density, repulsive
Coulomb interaction, temperature and bare diagonal hopping term $t^{\prime }$%
. All the energies are measured in units of $t$. In Figs.\~1 and 2 our
theoretical results for $n$ vs. $\mu $ are presented and compared with the
data obtained by a numerical study of a $8\times 8$ two-dimensional lattice%
\cite{Duffy:1995}. In Fig.\~3 the double occupancy $D$ is reported as
function of the particle density. As it can be seen a negative $t^{\prime }$
decreases $D$ when compared with $t^{\prime }=0 $, while a positive value
increases it. At half-filling the double occupancy is independent of the
sign of $t^{\prime }$ as it is required by the symmetries of the model, and
converges to the result for $t^{\prime }=0$. The agreement with the
experimental data given in Ref. \onlinecite{Duffy:1995} is generally quite
good.

\section{Conclusions}

\label{con}

By means of the $COM$, we have obtained a fully self-consistent solution for
the $t$-$t^{\prime }$-$U$ model. As for the simple Hubbard model, also in
the case of the $t$-$t^{\prime }$-$U$ model our scheme of calculation can
reproduce with good accuracy the results of numerical simulation. In a
forthcoming paper we shall continue the analysis of this model by
considering the magnetic and transport properties which characterize the
anomalous normal-state properties of the cuprates.

\end{document}